\documentclass[a4paper]{article}
\usepackage{graphicx}
\usepackage{amsfonts,amssymb,enumerate,epsf}
\usepackage{authblk}
\usepackage{float}
 \usepackage[english]{babel}
\DeclareMathSizes{11}{9}{7}{5}
 \def\botcaption#1#2{\medskip\centerline{{\scshape #1.}\kern8pt
 {\rm #2}}\bigskip}
\def \ba {\begin{array}}
\def \ea {\end{array}}
\def \be {\begin{equation}}
\def \ee {\end{equation}}

\def\bal{\begin{aligned}}
\def\eal{\end{aligned}}

 \def \R {{\mathbb R}}

\def \bg{{\mathbf g}}

\def \bk{{\mathbf k}}
\def \bx{{\mathbf x}}
\def \bu{{\mathbf u}}
\def \bv{{\mathbf v}}
\def \bY{{\mathbf Y}}

 \def \cO {{\cal O}}

 \def \ep {{\epsilon }}
\def \la {{\lambda}}
\def \La {{\Lambda}}

 \def \p {{'}} 

\def \un {{(1)}}

\def \du {{(2)}}

\def \zer {{ (0) }}
 \def \vet#1 {{\bf {#1}}}

\begin{document}
  \author{
 C. Boldrighini
 \footnote{
 Istituto Nazionale di Alta Matematica,
 Universit\`a di Roma Tre,
  Largo S.\ Leonardo Murialdo 1, 00146 Rome,
 Italy.
  }\\
 S. Frigio \footnote{ 
 Scuola di Scienze e Tecnologie, Universit\`a di Camerino, 62932 Camerino, Italy.  }\\ 
 
 P. Maponi \footnote{ 
 Scuola di Scienze e Tecnologie, Universit\`a di Camerino, 62932 Camerino, Italy.  }\\ 

   A. Pellegrinotti
 \footnote{
 Dipartimento di Matematica e Fisica,
 Universit\`a di Roma Tre,
 Largo S.\ Leonardo Murialdo 1, 00146 Rome,
 Italy. Partially supported by  research funds of INdAM (G.N.F.M.),  M.U.R.S.T. and Universit\`a Roma Tre.
 }\\
 Ya. G. Sinai \footnote { Dept. of Mathematics, Princeton University and Russian Academy of Sciences
 }}

 \title { AN ANTISYMMETRIC SOLUTION OF THE  3D INCOMPRESSIBLE  NAVIER-STOKES EQUATIONS }
 
 \maketitle
 
    \begin{abstract}
 We report results on the behavior of a particular  incompressible Navier-Stokes (NS) flow in the whole space $\R^{3}$,  related to the complex singular solutions introduced by  Li and Sinai in \cite{LiSi08} that blow up at a finite time.  The flow exhibits for some time a tornado-like behavior: a sharp  increase of vorticity and of the maximal velocity, with concentration in an annular region around an axis.  There is however no rotation around the axis and the approximate axial symmetry with no swirl excludes a real blow-up.  We conclude with a discussion on the possible candidates for a real blow-up.   \end{abstract}   \par \medskip
  
 Keywords:  Incompressible Navier-Stokes, Singular solutions, Blow-up, Tornadoes

   \section {Introduction}
  \label {P0}
We report results, obtained by computer simulations, guided by a theoretical analysis,  on a new solution of the  incompressible Navier-Stokes equations (NS) in $\R^{3}$, with no boundary conditions, which arises in connection with the contribution of Li and Sinai \cite{LiSi08} to the ``global regularity problem'',
    i.e., the problem whether smooth solutions in absence of forcing can become singular at a finite time. The problem is open since the pioneering work of J. Leray \cite{Leray} in 1937, and, in spite of many brilliant contributions, it is still open and  in the list of the Clay millennium prize.      
     Mathematically speaking the most notable difficulties of NS equations in 3D are its
non-locality, due to 
incompressibility, and super-criticality. The system is  super-critical with respect to the
basic energy conservation law.       \par\smallskip
     Leray, who first proved a global weak existence theorem and a uniqueness and regularity theorem only for finite times, believed that
there is loss of regularity and it  is related to  turbulence. The modern view of turbulence involves transitions to chaotic flows rather than  singularities. However, if singularities exist, they could describe   sudden concentrations of energy in a finite  region, as it happens in tornadoes or hurricanes,  for which no effective model is now available. In fact, the main features of the possible finite-time singularities (``blow-up''), are the divergence of the total enstrophy \cite{Temam},  and the divergence at some point
 of the absolute value of the velocity \cite{Seregin12}.\par\smallskip
 In recent times, results of  discrete models of the NS equations which preserve the energy conservation \cite{Ch08, KP02} seem to indicate that blow-up's are real. A blow-up was also proved  by T. Tao \cite{TT14}  for a continuous model obtained by modifying the bilinear term of the NS equations. (The introduction is a good review of the state of the global regularity problem.)   
As for the evidence  from computer simulations,   the NS equations are in general   difficult to follow on the computer, especially   for high values of the velocity and the vorticity, and in absence of reliable theoretical guide-lines on the structure of the blow-up the computer simulation are inconclusive  (see, e.g., \cite{Hou09})  
\par\smallskip

In 2008 a paper of Li and Sinai  \cite{LiSi08}   introduced a new approach. It is based on dynamical system techniques, which allow to  control, for some classes of flows, the tranfer of energy to the fine scales.    The  approach can  also be applied  to other models \cite{LiSi10(2)}. We give here a brief description. \par\smallskip
Consider the NSS in the whole space $\R^{3}$ with no forcing
\be\label{eq1} {\partial {\bf u}\over \partial t}(\bx,t) + \sum_{j=1}^3 u_j(\bx,t) {\partial {\bf u} \over \partial x_j}(\bx,t) =\nu \Delta \mathbf u(\bx,t) - \nabla p(\bx,t), \qquad  {\bf x} = (x_1, x_2, x_3)\in \R^3 .  \ee
   \be\label{12}\nabla \cdot \mathbf u(\bx,t) = \sum_{j} {\partial u_{j}\over \partial x_{j}}(\bx,t) = 0, \qquad  \mathbf u(\bx, 0) = \mathbf u_0(\bx), \ee
where $\nu$ is the kinematic viscosity and $p$ is the pressure, which can, by incompressibility,  be recovered in terms of the velocity field by a Biot-Savart law. Assuming $\nu=1$ (which can always be obtained by a suitable scaling), and introducing the modified Fourier transform
 \be\label{utilde} \mathbf {\mathbf v}(\mathbf k, t) = {i\over (2\pi)^{3}} \int_{\R^3} \mathbf u(\mathbf x, t) e^{-i \langle \mathbf k, \mathbf  x\rangle } d{\mathbf x}, \qquad \mathbf k= (k_1, k_2, k_{3}) \in \R^3,\ee
  where $\langle \cdot, \cdot \rangle$ denotes the  scalar product in $\R^3$, the equation (\ref{eq1}) can be written, by a Duhamel formula, as a single integral equation: 
   \be \label {kequat}  \mathbf v ({\bf k},t) = e^{-t {\bf k}^2} \mathbf v_{0}({\bk})  +\int_0^t  e^{-(t-s)|\bk|^2}\int_{\mathbb{R}^3}
\left \langle \bv(k-k',s), \bk\right \rangle P_{\bk} \bv(\bk',s) d\bk' ds,\ee 
where $P_{\bk} \bv = \bv - {\langle \bv, \bk \rangle \over |\bk|^{2}} \bk$ denotes the solenoidal projector and $\bv_{0}$  is the transform of  $\bu_{0}$.   In general  $\bv(\bk,t)$ is a complex function. Li and Sinai consider only real solutions of (\ref{kequat}), which in general correspond to complex solutions of (\ref{eq1}). However  if   $\bv_{0}(\bk)$ (and hence $\bv(\bk,t)$ for $t>0$)  is antisymmetric, the solution  $\bu(\bx,t)$ is also real and antisymmetric in $\bx$.  \par\smallskip
Multiplying the initial data by a real parameter $A$, which controls the initial energy, and
 iterating the Duhamel formula, the solution of (\ref{kequat}) is written as a power series:
 \begin{equation}\label{serie} \bv_{A}(\bk, t) =  A   \mathbf g^\un(\mathbf k, t)+ \int_0^t e^{-{\mathbf k}^2(t-s) } \sum_{p=2}^\infty A^p \mathbf g^{(p)}(\mathbf k, s) ds, \end{equation}
where  $\mathbf g^\un(\mathbf k, s) = e^{- s {\bk}^2} \bv_{0}(\bk)$,  $  \mathbf g^{(2)}(\mathbf k, s) =     \int_{\R^3} \left \langle \mathbf g^{\un}(\bk-\bk\p, s),\bk \right \rangle   P_{\bk} \mathbf g^{\un}(\bk\p,s)   d {\bk}\p $ and for $p>2$
\be\label{0}    \mathbf g^{(p)}(\mathbf k, s) 
     =  \sum_{p_1 + p_2 = p\atop p_1, p_2 >1} \int_0^s ds_1 \int_0^s ds_2 
   \int_{\R^3} \left  \langle \mathbf g^{(p_1)}(\mathbf k - \mathbf k\p, s_1), \bk \right \rangle \cdot $$ $$\cdot   P_{\bk}\mathbf g^{(p_2)} (\bk\p, s_2) e^{-  (s-s_1) (\bk - \bk\p)^2 - (s-s_2) ( \bk\p)^2} d\mathbf k\p  + {\rm boundary \; terms}. \ee
  The boundary terms involve $\mathbf g^{\un}$ and have a slightly different form (see, e.g.,\cite{BFMPS19}, where
 it is also shown, under some general conditions on $\bv_{0}$,  that   the series converges if $|A|t$ is small).

\par\smallskip

 Li and Sinai \cite{LiSi08} consider   real initial data $\bv_{0}$ such that  the support  is mainly concentrated inside  a sphere $K_{R}$ of radius  $R$ centered around the point $\bk^{\zer}$ with $|\bk^{\zer}| >> R$. Then   the function $\bg^{(p)}$, which is essentially a convolution,  has a support  centered around $p \bk^{\zer}$ and (by analogy with probability theory) an effective diameter of the order $\cO(\sqrt p)$.   \par\smallskip  
 Li and Sinai argue  that the  behavior of  $\mathbf g^{(p)}$ for large $p$ is determined by the fixed points of the map $\tilde \mathbf g^{(p)}\to  \tilde \mathbf g^{(p+1)}$, where $\tilde \mathbf g^{(p)}(\bY, s) = \mathbf g^{(p)}(p \bk^{\zer}+ \sqrt p \bY,s)$ are the functions  in the rescaled variables (rescaled as  for  the Central Limit Theorem). Hence, if we know the fixed point corresponding to the initial data we can control  the excitation of the high $\bk$-modes, i.e., of the fine structure components of $\bu(\bx,t)$.
 \par\smallskip
 More precisely, in \cite{LiSi08}  the following  {\it Ansatz} is formulated: for a class of gaussian dominated initial data with  support as described above, as $p\to\infty$ the following asymptotics holds
   \be \label{ansatz} \tilde \mathbf g^{(p)}(\bY,s) \sim p (\La(s))^{p}  \prod_{i=1}^{3 }g(Y_{i}) (\mathbf H(\bY) + \delta^{(p)}(\bY,s)), \qquad \bY = (Y_{1}, Y_{2}, Y_{3}) \ee where $\mathbf H$ is the fixed point,  $g(x) = {e^{-{x^{2}\over 2 }}\over \sqrt {2\pi} }$ is the standard Gaussian density, $\La$ is a strictly increasing smooth positive function and   $\delta^{(p)}(\bY,s)\to 0$  as $s\to \infty$.  (Assuming the standard gaussian is not restrictive, as it can   always be obtained by rescaling. )  The fixed point equation  for $\mathbf H$ \cite{LiSi08, BLS17, BFM17}  has infinitely many solutions that can be explicitly  written down \cite{LiSi08}. ($\mathbf H$ is in fact a plane vector, as its component along $\bk^{\zer}$ vanishes by incompressibility.)\par\smallskip
   
     Assuming  $\bk^{\zer}= (0,0,a)$, with $a>0$,      the   {\it Ansatz} (\ref{ansatz}) is proved in \cite{LiSi08} for the  fixed point    $\mathbf H^{\zer}(\bY) = c (Y_{1}, Y_{2}, 0)$, where $c$ is a real constant.   It turns out that the linearized map at the fixed point has a 6-dimensional unstable subspace and a 4-dimensional neutral subspace, in addition to an infinite-dimensional stable one. 
      The main
    result  of \cite{LiSi08} can be formulated  as follows. Let $a> b >> 1 $, and consider  initial data  of the form 
\be\label{iniziali} \bv_{0}(\bk) = \bar \bv(\bk) =  A  \left [   \left ( k_{1}, k_{2}, - {k_{1}^{2}+ k_{2}^{2} \over k_{3} }\right ) + \Phi(k_{1}, k_{2}, k_{3}) \right ] 
\prod_{i=1}^{2}     g(k_{i})   
g(k_{3}-a)     \chi_{b}(k_{3} - a), \ee 
      where  $\chi_{b}(k_{3})$ is  smooth and such that 
     $\chi_{b}(k_{3})=0$ if $|k_{3}|\geq b$,  $\chi_{b}(\bk)=1$ if $|\bk|\leq b-\ep$, with  $\ep  $  small enough,    $\Phi= \Phi^{\un}+ \Phi^{\du}$,   $\Phi^{\un}$ is a linear combination of the unstable  and neutral eigenfunctions of the linearized  map near $\mathbf H^{\zer}$, and $\Phi^{\du}$ is a vector in the stable subspace.   Then if $\Phi^{\du}$ is small enough,   there are a time interval $(S_{-} \leq s \leq S_{+})$ and  
         an open set of the parameters defining $\Phi^{\un}$ for which the {\it Ansatz} (\ref{ansatz}) holds.\par\smallskip
         The blow-up is an easy consequence: if $A= \pm { 1\over \La(\tau)}$,  $\tau \in (S_{-}, S_{+})$, the series  (\ref{serie})  diverges as $s\uparrow \tau$.
      Observe that both the total enstrophy and the total energy diverge as $t\uparrow \tau$ (for complex function the energy equality holds but it is not coercive).   
       \par\smallskip
 Coming back to real flows, it is natural to consider initial data obtained by antisymmetrizing the  data (\ref{iniziali})  associated to  solutions that blow-up.   Their support  is now concentrated in two  finite regions around  $\pm \bk^{\zer}$, and the convolution $\mathbf g^{(p)}$ is    a sum of terms centered around the points $(0,0, \ell a)$ with $\ell= -p, \ldots, p$, with the main contribution coming for $|\ell| = \cO(\sqrt p)$. We have again a simple mechanism which, as the components $\mathbf g^{(p)}$ are excited,  moves the support of the solution to the high $\bk$ region. It is weaker than for the complex case, but it is rather efficient, especially if $a$ is large. 
 \par\smallskip
 The present paper is a brief report on the behavior of one such solution, which exhibits some features of a ``tornado'', such as a sharp increase of the absolute values of velocity and vorticity in a confined region, but does not seem to blow up. As we argue below, this is possibly due to the axial symmetry of the fixed point $\mathbf H^{\zer}$.  We expect however that 
 the results of \cite{LiSi08} on the blow-up hold also for other fixed points $\mathbf H \neq \mathbf H^{\zer}$, which are in general not axial symmetric, and we are sure that. In absence of theoretical results, important information can be obtained on the complex flows and their related real flows by computer simulations, which can also reveal physically relevant details.
  
 \par\smallskip
From the mathematical  point of view one should extend the fixed point analysis to the real antisymmetric solutions, and also extend the whole theory to other fixed points $\mathbf H \neq \mathbf H^{\zer}$.
\par\smallskip
The plan of our paper is as follows. In \S 2 we give some results of computer simulations, and \S 3 is devoted to  concluding remarks.

   \section {Results of computer simulations}
  \label {P1}

 The initial data of the real flow are obtained by antisymmetrizing  the function (\ref{iniziali})  with  $\Phi\equiv 0$ and $A>0$ (the simulations show that   $\Phi$ does not have much influence on the behavior of the solution):   
  \be\label {datinr} \bv^{\zer}(\bk) = A \left (k_{1}, k_{2}, -{k_{1}^{2}+k_{2}^{2}\over k_{3}} \right ) g(k_{1}) g(k_{2}) \left [ g(k_{3}-a) \chi_{b}(k_{3}-a) + g(k_{3}+a) \chi_{b}(k_{3}+a) \right ].  \ee   \par\smallskip
  We used a special program for  solutions of the integral equation (\ref{kequat}), created for the purpose of following the blow-up of the complex solutions, as described in the paper  \cite{BFM17}, where solutions of (\ref{kequat}) with initial data of the type (\ref{iniziali})  could be  followed up to times close to the critical blow-up time. \par\smallskip
 Our mesh  in $\bk$-space is  a regular lattice centered at the origin with step $\delta=1$, with maximal configuration $[-254, 254] \times [-254, 254] \times [-3000, 3000]$. We deal with  about $5 \cdot 10^{9}$ real numbers, close to the maximal capacity of modern supercomputers. \par\smallskip
 As in the previous work \cite{BFM17},  our  aim was to follow the behavior of the total enstrophy and of its marginal distributions in $\bk$-space, which describes the flow of energy to the microscale  in physical space.  The mesh step $\delta=1$ allows  to follow  the solution for  values of $|\bk|$  of a few thousand. The behavior of the total enstrophy and its marginals in $\bk$-space is stable with respect to refinements of the mesh,  as it  is mainly due to the extension of the support  along the $k_{3}$-axis (the transversal diameter of the support grows more slowly).  On the other hand the mesh step  is   rough for the initial values (\ref{datinr}), which have a diameter of the essential support of the order of a few units, so that  the large scale behavior of the solution $\bu(\bx,t)$ is poorly reproduced, as it happens with  the axial symmetry (as it happens for the axial symmetry in some figures below).  An analysis comparing the accuracy of our program with respect to that of  finite-difference methods is under way.  \par\smallskip

We  report results obtained by simulating a particular solution  with initial data (\ref{datinr}): $a=30$ and $A$ is such that the initial energy $E_{0}= {1\over 2} \int_{\R^{3}} |\bu(\bx,t)|^{2} d\bx = 2.5 \times 10^{5}$. The study of the behavior of the solutions as the parameters $a$ and $A$ vary is under way. Recall however that the NS scaling holds:  If $\bv(\bk,t)$ is a solution of  (\ref{kequat}) with initial data $\bv_{0}(\bk)$, and $\la>0$   then the function $\bv^{(\lambda)}(\bk, t) = \la^{2}\bv(\la \bk, \la^{-2}t)$ is also a solution with initial data $\la^{2}\bv_{0}(\la \bk)$. 
\par\smallskip
In what follows time is measured in units of $\tau = 1.5625 \times 10^{-8}$. As for the complex case \cite{BFM17} the large initial data ensure a short running time, which makes simulations possible.\par\smallskip
The first remarkable feature of the flow is that the total enstrophy $S(t) = \int_{\R^{3}} |\omega(\bx,t)|^{2}d\bx$, where $\omega(\bx,t)$ is the vorticity field, increases sharply up to a critical time $T_{E}\approx 711 \tau$ and then decays. The same pattern is followed  by the maximal value of the speed $|\bu(\bx,t)|$, except that the critical value is $T_{V}\approx 400 \tau$.

  \begin{figure}[H]  
\centerline {
\includegraphics[width=3.5in]{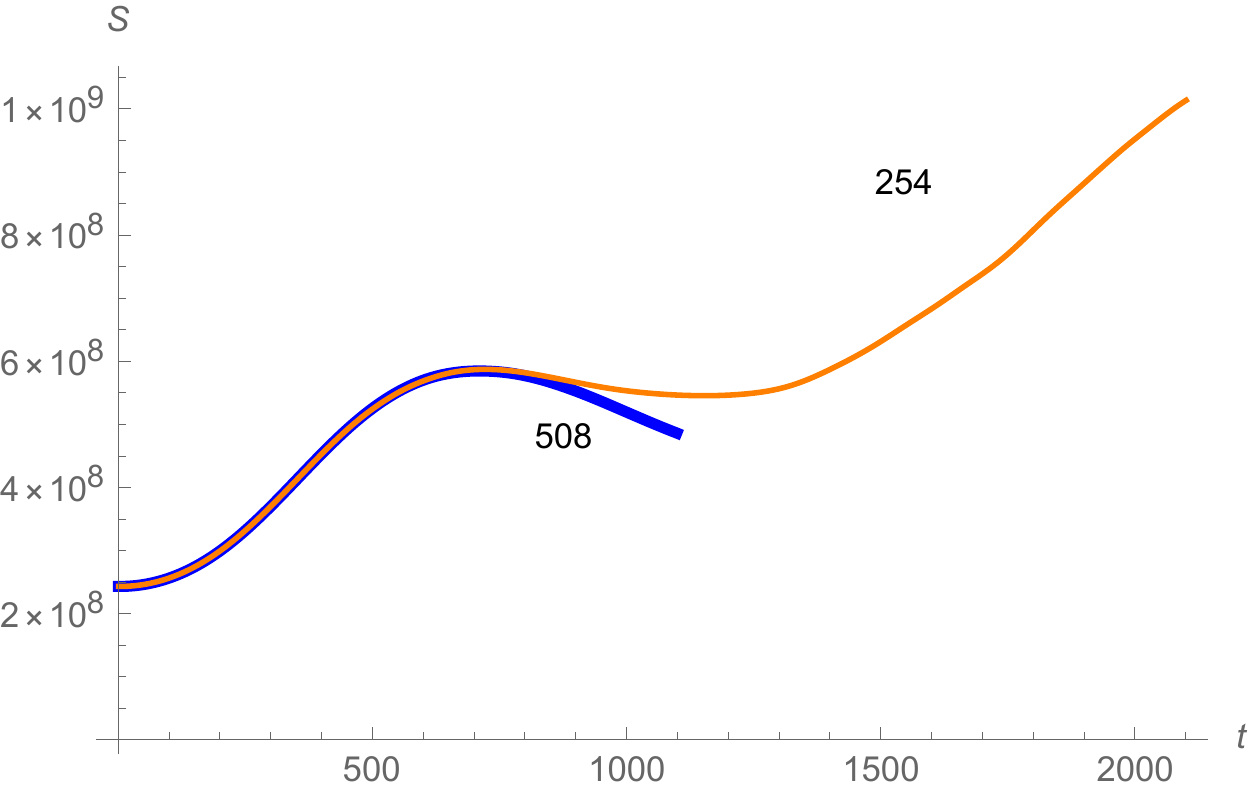}\hfill \includegraphics[width=3.5in]{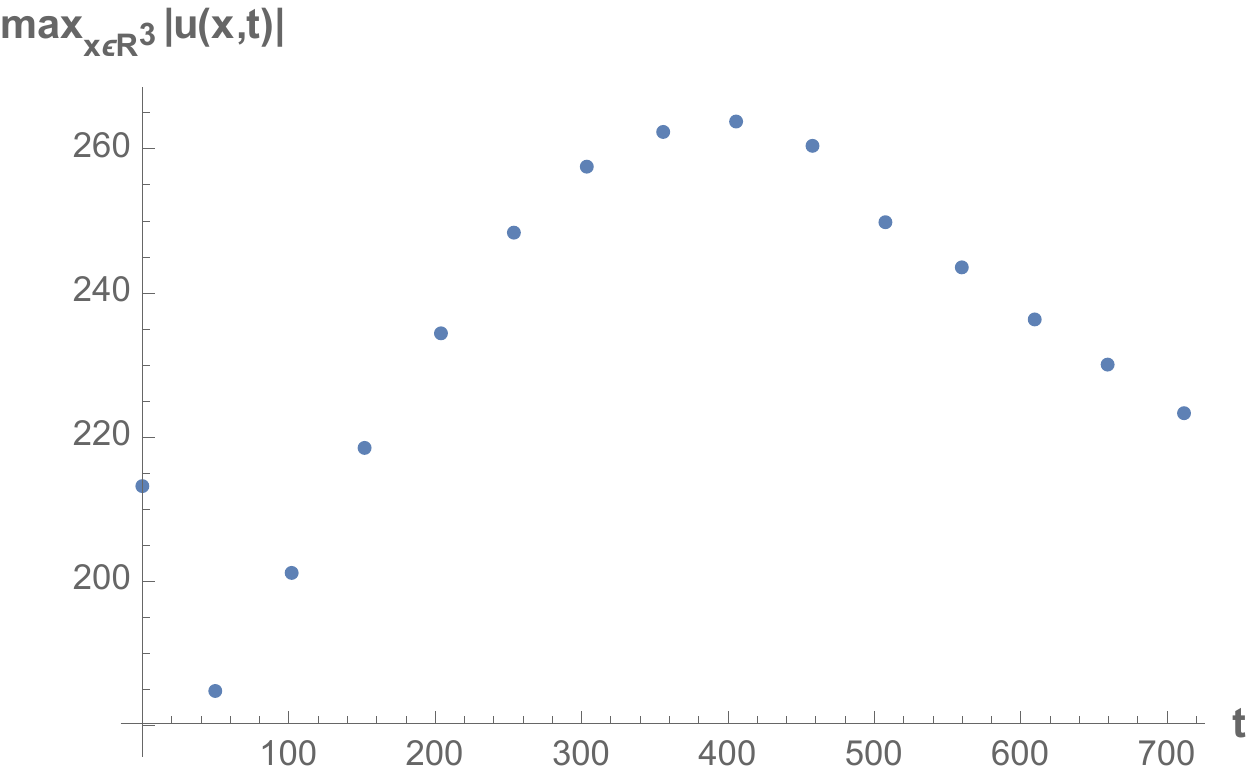} 
}  
 \caption{\it   Left side: Plot of the total  enstrophy $ S(t)$ with mesh $ [-147, 147]^{2} \times [-3000, 3000]$ (orange) and   mesh $ [-254, 254]^{2} \times [-3000, 3000]$ (blue). Right side: Plot of the maximal velocity as a function of time. }\label{Fig.1}\end{figure}
 
 In Fig. 1 (left) the blue line is a more precise description of the behavior of the total enstrophy. The orange line describes  a ``false blow-up''  due to a spurious production of enstrophy at the boundary corresponding to the extremal values $k_{1}, k_{2}= \pm 147$  of the narrow mesh. \par\smallskip
 Fig. 2 below describes the evolution in time of the marginal densities of the enstrophy along the third axis:  in $\bk$-space (left)   $S_{3}(k_{3},t) = \int_{\R^{2}} |\bk|^{2} |\bv(\bk,t)|^{2} dk_{1}dk_{2}$ (left), and in $\bx$-space $\tilde S_{3}(x_{3},t) = \int_{\R^{2}} |\omega(\bx,t)(\bx,t)|^{2} dx_{1}dx_{2}$. 
  \begin{figure}[H]  
 \centerline {
\includegraphics[width=3.5in]{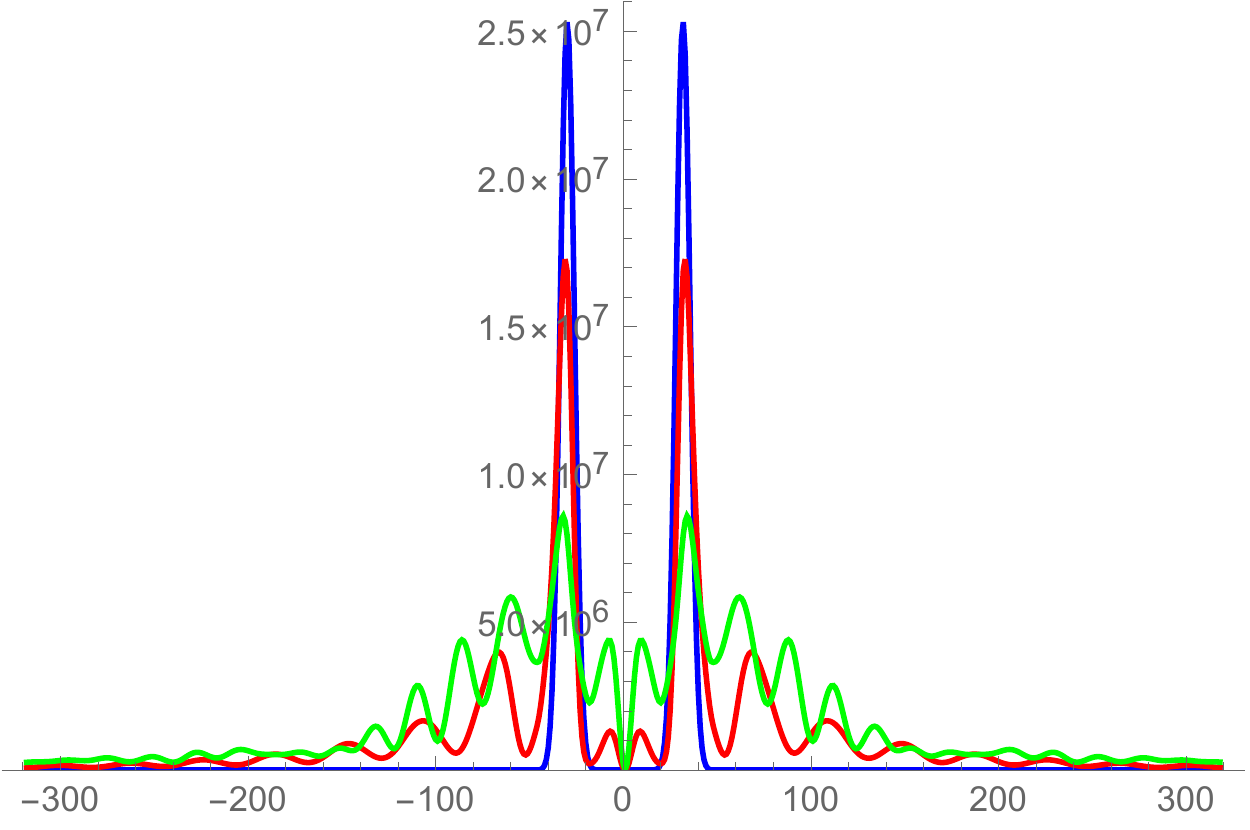}\hfill \includegraphics[width=3.5in]{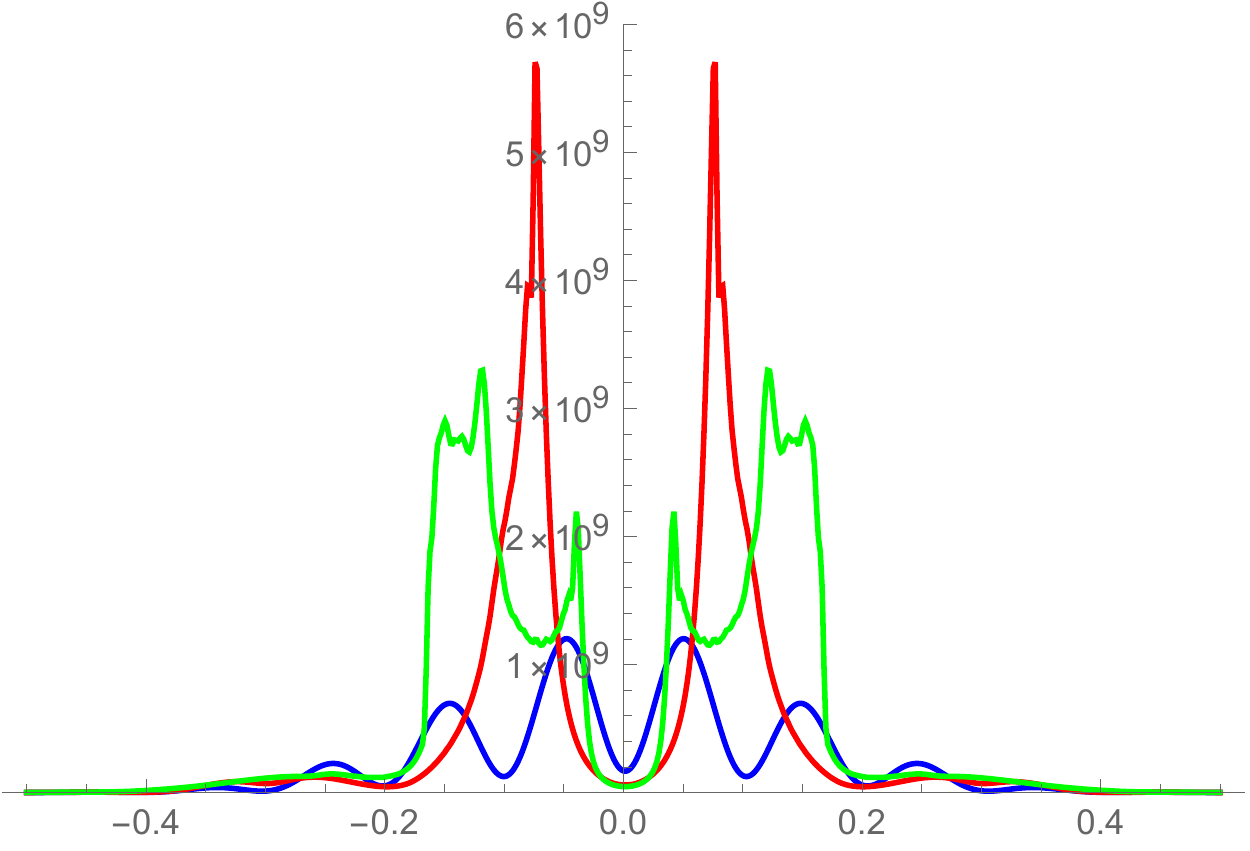} 
}  
 \caption{\it   Plots of the marginal distributions $S_{3}(k_{3},t)$ (left) and $\tilde S_{3}(x_{3},t)$ (right) at the times  $t=0$ (blue),  $t=400\tau \approx T_{V} $ (red), and $t=711\tau\approx T_{M}$ (green).    }\label{Fig.2}\end{figure}
 
On the left side of Fig.2  one can see how the support of the solution moves into the high $|\bk|$-region. As predicted by the theory, as time grows  the  peaks of the figure on the left tend to be close to the values $k_{3}\approx j a$, with $j= \pm1, \pm2, \ldots$ (green line), a modulated periodicity corresponding in the figure on the right to the peaks of the green line at  $x_{3} = \pm \bar x_{3}$ with $\bar x_{3}\approx {\pi\over a}$.   
\par\smallskip
The evolution of $\tilde S_{3}(x_{3},t)$ as illustrated by Fig. 2  also shows that the vorticity is concentrated in the neighborhood of symmetric pairs of planes orthogonal to the $x_{3}$-axis.  This is also true for the large values of the velocity field $\bu(\bx,t)$, as shown by the following Fig. 3, where we report the absolute value of the velocity field on the plane $x_{3}= 0.08$, corresponding to the right red peak on $\tilde S_{3}(x_{3}, 400)$ of Fig. 2.  

 \begin{figure}[H]  
  \centerline {
\includegraphics[width=3.5in]{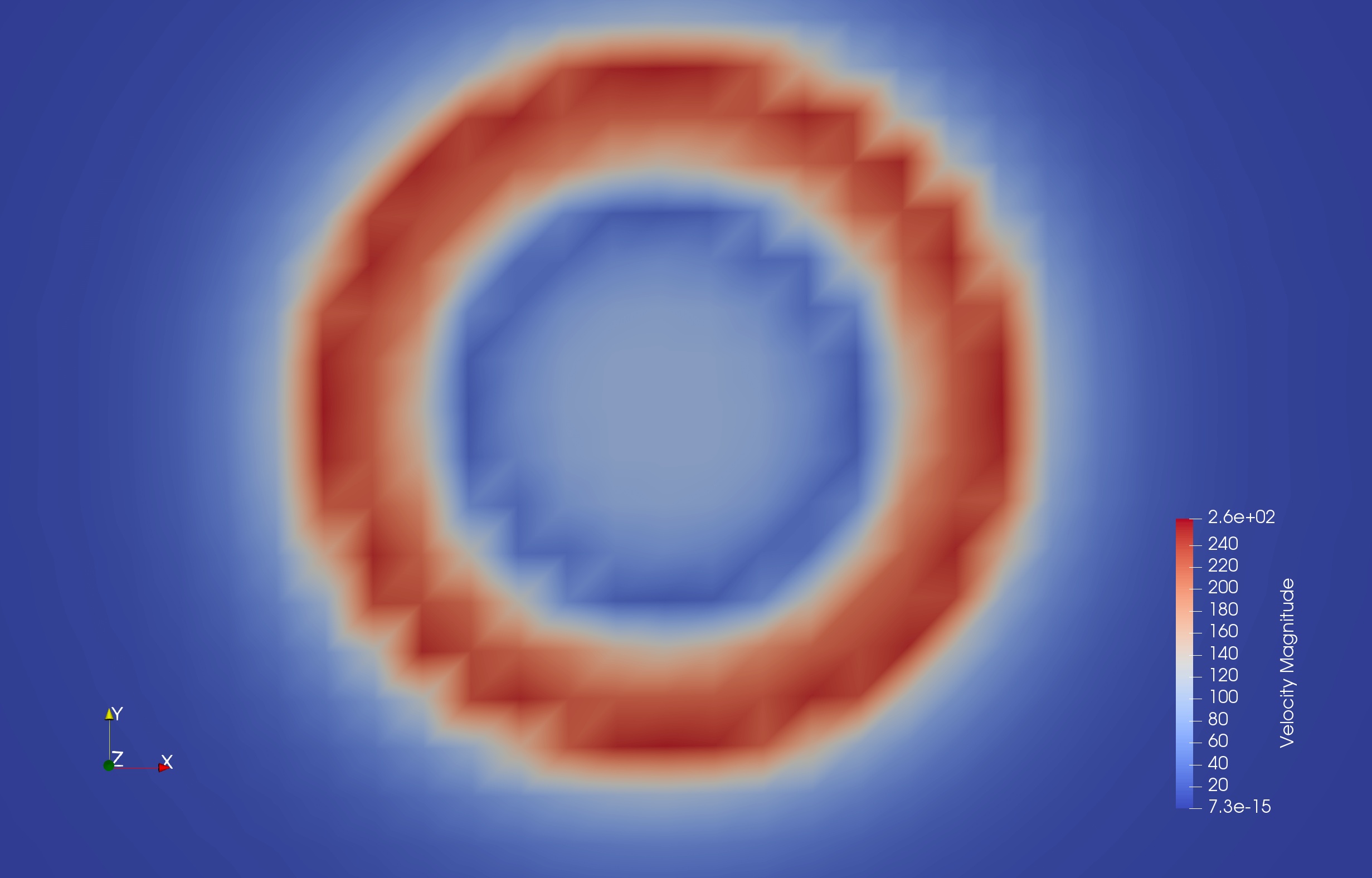}
}  
\caption{\it   The absolute value velocity  field $|\bu(\bx,t)|$ on the plane $x_{3}=0.03$ at the time $t=400\tau$. The red color indicates the highest values. }\label{Fig.3}\end{figure}
 \par\smallskip
Much interest has been  devoted in recent times to the behavior of the angle between the velocity $\bu(\bx,t)$ and the vorticity $\omega(\bx,t) = \nabla \times \bu(\bx,t)$ \cite{Be12, FG19}. In our case the average cosine of the angle, weighted with the local energy $|\bu(\bx,t)|^{2}\over 2$, is zero at the initial time  (orthogonality) and increases up to the time $T_{E}$, when it reaches a value close to $0.45$. The solution could not be followed after time $T_{E}$ with enough accuracy (as it appears, it begins to decrease), so that we have no hint on whether ``beltramization'' \cite{FG19} is taking place.

\par\smallskip

   \section {Concluding remarks}
  \label {P2}
 
As shown by Figg. 1, 2, 3, up to the critical time $T_{V}\approx 400 \tau$ the maximal velocity has a significant increase, and the vorticity  also grows and undergoes a sharp concentration in two ``doughnuts'' around the $x_{3}$-axis, bisected by the planes $x_{3}\approx \pm 0.8$. The velocities are also maximal in the ``doughnuts'',  while elsewhere the fluid stays quiet.  The total enstrophy increases up to the critical time     $T_{M}\approx 711 \tau$, while the maximal velocity decreases, and the sharp concentration persists, although the ``doughnuts'' undergo some deformation.  \par\smallskip

 The description is strongly remindful of tornadoes and hurricanes, except perhaps that in our case there is no rotation around the $x_{3}$-axis.  In fact the solution with initial data (\ref{datinr}) is axial-symmetric with no swirl, although the axial symmetry is partially lost in Fig. 3 due to computation errors, as we explained above.

 \par\smallskip
 
 Concerning the global regularity problem, it is well-known that regularity for all times holds for axial symmetric solutions with no swirl and for solutions close to them \cite{LZ17}. As the fixed point $\mathbf H^{\zer}$ is axial symmetric, the possible violations of symmetry and/or the initial swirl which can be introduced by inserting in the initial data (\ref{datinr}) the function $\Phi$ in (\ref{iniziali}) would not produce a singularity. As we said above, it is  more promising to look at the real flows associated, again by antisymmetrization, to initial data related to fixed points $\mathbf H \neq \mathbf H^{\zer}$   which are not axial symmetric. This will be the object of further work.  \par\bigskip

 \noindent {\bf Acknowledgements.}  The computer simulations were performed at the Marconi Supercomputer of CINECA (Bologna, Italy), within the framework of a European PRACE Project n. 2015133169, and also of CINECA ISCRA Projects of type B and C.

 \end{document}